\documentclass[11pt, a4paper]{article}
\usepackage{jheparxiv}
\usepackage[utf8]{inputenc}
\usepackage{amsmath}
\usepackage{amsfonts}
\usepackage{amssymb}
\usepackage{latexsym}
\usepackage{mathrsfs}
\usepackage{braket}		
\usepackage{graphicx}
\usepackage{color}
\usepackage{xcolor}
\usepackage{slashed}
\usepackage{twistor}
\usepackage{tensor}
\usepackage[all]{xy}

\newcommand{\sa}{\mathsf{a}}

\newcommand{\veps}{\varepsilon}
\newcommand{\eps}{\epsilon}

\renewcommand{\d}{\mathrm{d}}

\subheader{\hfill \texttt{IMPERIAL-TP-TA-2019-02}}

\title{Celestial amplitudes and conformal soft theorems}

\author[a]{Tim Adamo,}
\author[b]{Lionel Mason}
\author[b]{\& Atul Sharma}

\affiliation[a]{Theoretical Physics Group, Blackett Laboratory \\
        Imperial College London, SW7 2AZ, United Kingdom}
\affiliation[b]{The Mathematical Institute \\
        University of Oxford, Woodstock Road, OX2 6GG, United Kingdom}        

\emailAdd{t.adamo@imperial.ac.uk}
\emailAdd{lmason@maths.ox.ac.uk}
\emailAdd{atul.sharma@maths.ox.ac.uk}

\abstract{Scattering amplitudes in $d+2$ dimensions can be expressed in terms of a conformal basis, for which the S-matrix behaves as a CFT correlation function on the celestial $d$-sphere. We explain how compact expressions for the full tree-level S-matrix of gauge theory, gravity and other QFTs extend to this conformal basis, and are easily derived from ambitwistor strings. Using these formulae and their worldsheet origins, we prove various tree-level `conformal soft theorems' in gauge theory and gravity in any dimension; these arise from limits where the scaling dimension of an external state in the scattering process takes special values. These conformally soft limits are obscure from standard methods, but they are easily derived with ambitwistor strings. Additionally, we make an identification between the residues of conformally soft vertex operator insertions in ambitwistor strings and charges generating asymptotic symmetries.}  

\begin{document}
\notoc

\maketitle\vfill
 
\section{Introduction}

The external legs in a massless scattering process are solutions of the free equations of motion, often expressed in a plane wave momentum eigenstate basis. In this plane wave basis, external states are specified by a momentum $p^\mu$ obeying $p^2=0$ (and other relevant quantum numbers like polarization or charge). Recently, it was noted that there is another basis which represents massless spin $s$ free fields in $d+2$ dimensional Minkowski space as spin $s$ conformal primaries on the $d$-sphere~\cite{Pasterski:2016qvg,Pasterski:2017kqt,Pasterski:2017ylz}. The data of an on-shell momentum vector is traded for a point $k\in S^d$ and a (freely chosen) scaling dimension $\Delta$. 

Scattering amplitudes expressed in this \emph{conformal basis} are equivalent to a Mellin transform of the amplitudes in the standard momentum eigenstate basis. The resulting S-matrix elements transform as conformal correlators on $S^d$, with scaling dimensions of the insertions given by the choices of $\Delta$ for the external states. Viewing $S^d$ as the celestial sphere of null generators of $\scri$ -- the conformal boundary of $d+2$-dimensional Minkowski space -- S-matrix elements in the conformal basis are referred to as \emph{celestial amplitudes}. Since they manifest an underlying boundary conformal symmetry for bulk observables, celestial amplitudes are of interest as objects of study in attempts to understand holography for asymptotically flat space-times (cf., \cite{deBoer:2003vf,Kapec:2014opa,He:2015zea,Kapec:2016jld,Cheung:2016iub,Cardona:2017keg,Lam:2017ofc,Banerjee:2017jeg,Schreiber:2017jsr,Banerjee:2018gce,Stieberger:2018edy,Stieberger:2018onx}).

\smallskip

Celestial amplitudes manifest the conformal covariance of the massless S-matrix on the asymptotic $d$ sphere of null directions, but other properties of scattering amplitudes familiar from the momentum basis become obscure in the conformal basis. An important example is the soft physics of massless particles: S-matrix elements for gluons or gravitons where one particle becomes soft are constrained by soft theorems, giving rise to universal limits~\cite{Low:1954kd,Low:1958sn,Weinberg:1965nx,White:2011yy,Cachazo:2014fwa,Casali:2014xpa}. In the conformal basis, there is no notion of energy or frequency; for special values of the scaling dimension $\Delta$, gluons or gravitons become pure gauge but the celestial amplitudes nevertheless have nontrivial limits~\cite{Pasterski:2017kqt,Donnay:2018neh,Banerjee:2019aoy}. 

Frequency $\omega$ in the momentum basis and scaling dimension $\Delta$ in the conformal basis are related non-locally via a Mellin transform, so the behaviour of S-matrix elements in a `conformally soft limit' (i.e., when $\Delta$ approaches a pure gauge configuration) is not immediately clear. Recently, the structure of the conformally soft limit $\Delta\rightarrow 1$ was investigated for tree-level gluon scattering in four space-time dimensions~\cite{Fan:2019emx,Pate:2019mfs,Nandan:2019jas}. Through explicit calculations at low particle number and for specific helicity configurations, a four-dimensional conformal soft theorem was conjectured whose structure is essentially inherited from the energetic soft theorem. However, calculations for general particle number and polarization or for general dimension and other theories remain lacking.

\smallskip

In this paper, we present formulae for all massless tree-level celestial amplitudes and derive a variety of tree-level conformal soft theorems in gauge theory and gravity in any space-time dimension. The key tool which enables these calculations is \emph{ambitwistor string theory}, a class of chiral, constrained worldsheet models which describe massless QFTs~\cite{Mason:2013sva}. Ambitwistor strings underpin the scattering equations and Cachazo-He-Yuan (CHY) formulae~\cite{Cachazo:2013hca,Cachazo:2014xea} for massless scattering in the momentum basis~\cite{Mason:2013sva,Adamo:2013tsa,Casali:2015vta}, but they also encode the full non-linearity of the underlying classical equations of motion~\cite{Adamo:2014wea,Adamo:2018hzd} and their spectrum always imposes the free equations of motion~\cite{Berkovits:2018jvm,Adamo:2018ege}. Thus, they can equally well describe scattering in the conformal basis, leading to CHY formulae for celestial scattering amplitudes. Ambitwistor strings and CHY formulae have been used to study energetic soft limits and their connection to asymptotic symmetries~\cite{Schwab:2014xua,Adamo:2014yya,Geyer:2014lca,Lipstein:2015rxa,Adamo:2015fwa,Cardona:2017keg}, so it is no surprise that they are also useful in the conformal basis.

In section~\ref{CHY}, we briefly review the conformal basis and ambitwistor strings before deriving CHY formulae for celestial scattering in any space-time dimension. While our focus is on gauge theory and gravity, the general form of a tree-level $n$-point celestial amplitude in CHY form is shown to be:
\begin{multline}\label{caCHY}
A_n = \lim_{\veps\rightarrow 0^+}\prod_{i=1}^n\int_0^\infty\frac{\d t_i}{t_i}\,t_i^{\Delta_i}\,\e^{-\veps t_i}\,\delta^{d+2}\!\left(\sum_{i=1}^n \alpha_i\,t_i k_i\right) \int \d \mu_{n}\,\sideset{}{'}\prod_{j}\delta\!\left(\sum_{l\neq j}\frac{\alpha_j\alpha_lt_jt_l\,k_j\cdot k_{l}}{\sigma_{j}-\sigma_l}\right) \\
\times\, \mathcal{I}_{n}(\sigma_i, \alpha_it_ik_i,\ldots)\,,
\end{multline}
where $\{\Delta_i\}$ are the scaling dimensions of the external states; $\{k_i\}$ are insertion points on the sphere $S^d$; $\{\alpha_i=\pm1\}$ indicate whether each state is outgoing $(+)$ or incoming $(-)$; and $\veps\rightarrow 0^+$ is a regulator to ensure convergence of the integrals over $\{t_i\}$. Integrals over points on the Riemann sphere $\{\sigma_i\}$ are performed against a M\"obius-invariant measure $\d\mu_n$, and localized by the conformal basis scattering equations appearing in the M\"obius-invariant product of delta functions. The space-time theory is specified by the choice of CHY integrand $\mathcal{I}_n$, whose functional dependence on the quantum numbers differs from the momentum basis only by the replacement $p_i\rightarrow \alpha_it_ik_i$ for on-shell momenta.

Conformally soft limits are then investigated in section~\ref{CSLs}. Using the behaviour of ambitwistor string vertex operators in the relevant limits, we obtain several conformal soft theorems at tree-level in both gauge theory and gravity. For a colour-ordered Yang-Mills celestial amplitude, we find analytic continuations in $\Delta$ so that
\be\label{glcst1}
\lim_{\Delta\rightarrow 1}(\Delta-1)\,A_{n+1}(\Delta_1,\ldots,\Delta_n,\Delta)=\left(\frac{\eps\cdot k_1}{k\cdot k_1}-\frac{\eps\cdot k_n}{k\cdot k_n}\right)\,A_{n}(\Delta_1,\ldots,\Delta_n),
\ee
where $k\in S^d$ and $\epsilon_{\mu}$ are the insertion point and polarization of the $\Delta\rightarrow 1$ gluon. For tree-level celestial graviton amplitudes, we find conformal soft theorems for both $\Delta\rightarrow 1$ and $\Delta\rightarrow 0$:
\be\label{grcst1}
\lim_{\Delta\rightarrow 1}(\Delta-1)\,A_{n+1}(\Delta_1,\ldots,\Delta_n,\Delta) =\sum_{i=1}^{n}\alpha_i\,\frac{\eps_{\mu\nu}\, k^{\mu}_i\, k^{\nu}_i}{k\cdot k_i}\,A_{n}(\Delta_1,\ldots,\Delta_i+1,\ldots,\Delta_n)\,,
\ee
and
\be\label{grcst0}
\lim_{\Delta\rightarrow 0}\Delta\,A_{n+1}(\Delta_1,\ldots,\Delta_n,\Delta) =\sum_{i=1}^{n}\eps_{\mu\sigma}\,k_{\nu}\frac{k^{\sigma}_i\,J_{i}^{\mu\nu}}{k\cdot k_i}\,A_{n}(\Delta_1,\ldots,\Delta_n)\,,
\ee
for $J_i^{\mu\nu}$ the angular momentum operator associated to graviton $i$. We also give refinements of these conformal soft theorems in four space-time dimensions, with their derivation from ambitwistor strings in appendix~\ref{4D}.

The structure of \eqref{glcst1} -- \eqref{grcst0} is clearly related to that of leading and sub-leading energetic soft theorems (cf., \cite{Strominger:2017zoo}). From the perspective of ambitwistor strings, the reason for this is clear: the structure of a conformally soft vertex operator is identical to that of an energetically soft vertex operator. In section~\ref{asymptotics}, we use this fact to prove that conformally soft gluon or graviton vertex operators have the structure of charges which generate asymptotic symmetries of gauge theory and gravity at $\scri$. 

We also comment on a peculiarity in gauge theory: while $\Delta\rightarrow 0$ does \emph{not} correspond to a pure gauge gluon conformal primary, this limit does generate factorization behaviour at the level of the tree-level celestial gluon amplitudes, and produces an asymptotic symmetry associated with the sub-leading energetic soft theorem of Yang-Mills theory.


\section{CHY Formulae for Celestial Amplitudes}
\label{CHY}

The remarkable formulae of Cachazo-He-Yuan (CHY) for tree-level scattering amplitudes in a wide variety of massless QFTs take the universal form~\cite{Cachazo:2013hca,Cachazo:2014xea}:
\be\label{CHY0}
\cA_{n}=\delta^{D}\!\left(\sum_{i=1}^{n}p_i\right) \int \d \mu_{n}\,\sideset{}{'}\prod_{i} \delta\!\left(\sum_{j\neq i} \frac{p_i\cdot p_j}{\sigma_i-\sigma_j}\right)\, \mathcal{I}_{n}(\sigma_i, p_i,\ldots)\,,
\ee
where $D$ is the space-time dimension, $\{p_i\}$ are the on-shell $D$-momenta associated with the representation of each external state in a plane wave basis, and the integral is over the location of $n$ marked points $\{\sigma_i\}$ on the Riemann sphere with respect to a M\"obius-invariant measure $\d\mu_{n}$. These integrals are localized by the \emph{scattering equations},
\be\label{kSEqs}
\sum_{j\neq i} \frac{p_i\cdot p_j}{\sigma_i-\sigma_j}=0\,, \quad \forall i=1,\ldots,n\,.
\ee
Only $(n-3)$ of these equations are independent; the prime appearing on the product in \eqref{CHY0} ensures that only $(n-3)$ of these constraints are imposed, compatibly with the M\"obius-invariant measure (cf., \cite{Cachazo:2013gna,Dolan:2014ega}). The final ingredient in the CHY formula \eqref{CHY0} is the integrand $\mathcal{I}_{n}$, which is a rational function of the $\{\sigma_i\}$ and any quantum numbers associated to the scattering process (e.g., momenta, polarizations, colour charges); the choice of this integrand determines which theory's scattering amplitudes are described in this way. To date, integrands are known for gauge theory, gravity and many other massless QFTs in any dimension~\cite{Cachazo:2014xea}.

\emph{A priori}, it seems that the CHY formulae are closely tied to expressing amplitudes in a plane wave basis: \eqref{CHY0} is only well-defined on the support of overall momentum conservation, and the scattering equations \eqref{kSEqs} explicitly depend on the momenta associated to each particle. However, the worldsheet origin of the CHY formulae makes it easy to see that compact expressions for the tree-level S-matrix -- and the scattering equations -- extend to other choices of basis, including the conformal basis.


\subsection{The conformal basis}

From now on, we will take the space-time dimension to be $D=d+2$, working on Minkowski space $\R^{1,d+1}$ with signature $(-,+,\cdots,+)$. A massless scalar field $\phi$, satisfying
\be\label{scal1}
\Box\phi=0\,,
\ee
for $\Box$ the wave operator on $\R^{1,d+1}$, is usually represented in terms of a basis of Fourier mode wavepackets. An element of this basis is the familiar plane wave momentum eigenstate: $\phi^{\pm}(X) =\e^{\pm\im p\cdot X}$ for $p^2=0$ and $p^0>0$, the choice of sign dictating whether the state is outgoing $(+)$ or incoming $(-)$. 

A general analogue of Fourier transform enables a formal expansion via the inverse Mellin transform~\cite{Strichartz:1973}
\be\label{scal2}
\e^{\pm\im\, p\cdot X} = \int_{-\infty}^{+\infty}\frac{\d\nu}{2\,\pi}\, \frac{(\mp\im)^{h+\im\nu}\,\Gamma(h+\im\nu)}{(-p\cdot X)^{h+\im\nu}}\,,
\end{equation}
where $h$ is any positive real number. This expands  momentum eigenstates in homogeneous functions of $p$. In~\cite{Pasterski:2017kqt}, it was realized that this relation can be inverted to construct a new conformally covariant, delta-function normalizable basis for massless free fields $\phi(X)$ in terms of bulk-to-boundary propagators $(-p\cdot X)^{-(h+\im\nu)}$ in AdS. Now, $S^d$ naturally embeds into $\R^{1,d+1}$ as the projective light cone (i.e., the celestial sphere) via
\be\label{embedding}
    k:S^d\rightarrow\R^{1,d+1},\qquad k^\mu(x^a) = \left(\frac{1+x^2}{2},\,x^a,\,\frac{1-x^2}{2}\right)\,,
\ee
with $x^a$ flat coordinates on $S^d$. Solutions to the free field equation \eqref{scal1} can be constructed by continuing a bulk-to-boundary propagator to $\R^{1,d+1}$:  
\be\label{sCB}
    \phi^{\pm}_{\Delta}(X;k) = \int_0^\infty\frac{\d t}{t}\, t^\Delta\,\e^{\pm\im\, tk\cdot X-\veps t}= \frac{(\mp\im)^\Delta\,\Gamma(\Delta)}{(-k\cdot X\mp\im\veps)^\Delta},
\ee
where $\veps\rightarrow 0^+$ is a regulator, $\Delta$ is a complex number and $k\in S^d$. The sign $\pm$ again denotes outgoing/incoming states ($k$ is always future pointing).

Note that $\phi^{\pm}_{\Delta}$ solves the massless wave equation for \emph{any} value of $\Delta$; the role of $\Delta$ is to set a scaling dimension for the solution associated to a conformal transformation on $S^d$. A conformal transformation is given by the simultaneous group action
\begin{equation}\label{conftrans}
    \Lambda\in \mathrm{SO}(1,d+1),\qquad X^\mu\mapsto X'^\mu=\tensor{\Lambda}{^\mu_\nu}X^\nu,\quad k^\mu(x)\mapsto k'^\mu(x')=\Omega(x)^\frac{1}{d}\tensor{\Lambda}{^\mu_\nu}k^\nu(x),
\end{equation}
where $\displaystyle\Omega = \left|\frac{\partial x'}{\partial x}\right|$ is the Jacobian of the induced conformal transformation on $S^d$. Under such a transformation, the new state \eqref{sCB} transforms as
\begin{equation}
    \phi_\Delta^{\pm}(X;k)\mapsto\Omega(x)^{-\frac{\Delta}{d}}\,\phi_\Delta^{\pm}(X;k),
\end{equation}
and hence is a conformal primary on $S^d$ of scaling dimension $\Delta$.

In~\cite{Pasterski:2017kqt}, it was shown that a basis of these modes which are delta function normalizable in the Klein-Gordon norm and span the space of plane wave modes is given by:
\be\label{sbasis}
    \left\{\phi^\pm_{\Delta}(X;k)\left|\Delta\in \frac{d}{2}+\im\R,\: k^2=0,\,k^0+k^{d+1}=1\right.\right\}.
\end{equation}
The conditions on $k^{\mu}$ ensure that it labels a point on the sphere $S^d$, while the condition that $\Delta=\frac{d}{2}+\im\R$ is a consequence of harmonic analysis on the conformal group SO$(1,d+1)$, commonly known in the literature as the principal continuous series. The basis of states \eqref{sbasis} is known as the \emph{conformal basis}.

\medskip

Conformal bases have also been constructed for gluons and gravitons propagating on $\R^{1,d+1}$~\cite{Pasterski:2017kqt}. For gluons, the appropriate wavefunction is a solution to the Maxwell equations on $\R^{1,d+1}$ which transforms as a vector in $d+2$ dimensions and a spin-1 conformal primary on $S^d$ under SO$(1,d+1)$. A solution which satisfies both Lorenz ($\partial^{\mu}A_{\mu}=0$) and radial ($X^{\mu}A_{\mu}=0$) gauge conditions is given by:
\be\label{glCB1}
    A_{\mu}^{\Delta,\pm}(X;k) = (\mp\im)^\Delta\,\Gamma(\Delta)\,\frac{(\eps\cdot X)\,k_\mu-(k\cdot X)\,\eps_\mu}{(-k\cdot X\mp\im\veps)^{\Delta+1}}\,,
\ee
which is closely related to the spin-1 bulk-to-boundary propagator in AdS (cf., \cite{Costa:2014kfa}). Once again, $\Delta$ labels the scaling dimension and $k\in S^d$ is a point on the celestial sphere; the polarization vector takes the form:
\be\label{pol1}
    \epsilon_{\mu}:=\epsilon^{a}\,\frac{\partial}{\partial x^a}k_\mu(x)\,,
\end{equation}
for $k_{\mu}(x)$ given by \eqref{embedding}. The constants $\epsilon^a$ parameterize the $d$ degrees of freedom of an on-shell gluon in $\R^{1,d+1}$, and the condition $k^2=0$ ensures that $k\cdot\epsilon=0$.

A simpler solution to the Maxwell equations can be obtained from \eqref{glCB1} by a gauge transformation, leading to (up to normalization constants):
\be\label{glCB2}
    A_\mu^{\Delta,\pm}(X;k)= (\mp\im)^\Delta\,\Gamma(\Delta)\,\frac{\eps_\mu}{(-k\cdot X\mp\im\veps)^{\Delta}} = \eps_\mu\,\phi_\Delta^\pm(X;k)\,.
\ee
While this gauge transformed solution is no longer a spin-1 conformal primary (since $\eps_\mu$ depends on $k$ and transforms non-trivially), gauge-invariant observables like the S-matrix are unaffected. As for the massless scalar, a normalizable and complete conformal basis of gluon states is given by $A_{\mu}^{\Delta,\pm}(X;k)$ with $\Delta=\frac{d}{2}+\im\R$.

For gravitons the appropriate wavefunction is a solution of the linearised Einstein equations on $\R^{1,d+1}$ which transforms as a rank-2 symmetric tensor in $d+2$ dimensions and a spin-2 conformal primary on $S^d$ under SO$(1,d+1)$:
\be\label{grCB1}
h_{\mu\nu}^{\Delta,\pm}(X;k)= (\mp\im)^\Delta\,\Gamma(\Delta)\,\frac{\left[(\eps\cdot X)\,k_{(\mu}-(k\cdot X)\,\eps_{(\mu}\right]\left[(\tilde\eps\cdot X)\,k_{\nu)}-(k\cdot X)\,\tilde\eps_{\nu)}\right]}{(-k\cdot X\mp\im\veps)^{\Delta+2}}\,,
\ee
where $\tilde{\epsilon}_{\mu}$ is another copy of the polarization \eqref{pol1}. Once again, this can be simplified by using a diffeomorphism to arrive at:
\be\label{grCB2}
h_{\mu\nu}^{\Delta,\pm}(X;k)=(\mp\im)^\Delta\,\Gamma(\Delta)\,\frac{\eps_{(\mu}\,\tilde\eps_{\nu)}}{(-k\cdot X\mp\im\veps)^{\Delta}} = \eps_{(\mu}\,\tilde\eps_{\nu)}\,\phi_{\Delta}^\pm(X;k)\,.
\ee
While \eqref{grCB2} is not a spin-2 conformal primary, it is gauge-equivalent to \eqref{grCB1}. A normalizable and complete conformal basis of graviton states corresponds to $\Delta=\frac{d}{2}+\im\R$.

\medskip

So a conformal basis state is labeled by a point on $S^d$ and a scaling dimension $\Delta$ (along with the relevant polarization data), while a plane wave state is labeled by an on-shell momentum in $\R^{1,d+1}$. The conformal basis spans the same set of free fields as the plane wave basis, so the conformal basis is an equally valid one to use when amputating external legs in a massless scattering process via the LSZ procedure. A scattering amplitude evaluated on these states is referred to as a \emph{celestial amplitude} since the external states are specified by points on the celestial sphere $S^d$~\cite{Pasterski:2017kqt,Pasterski:2017ylz}. Celestial amplitudes can be obtained directly from the momentum basis by Mellin transforms
\be\label{CBAmps}
    A_n(\alpha_i,\Delta_i,k_i) = \prod_{j=1}^n\int_0^\infty\frac{\d t_j}{t_j}\,t_j^{\Delta_j}\,\e^{-\veps t_j}\,\cA_n(\alpha_it_ik_i),
\ee
in the limit $\veps\rightarrow 0^+$. Here, $\alpha_i=\pm 1$ represents whether the $i^\text{th}$-particle is incoming $(-)$ or outgoing $(+)$. From the definitions of the conformal states, it is clear that the conformal basis amplitude $A_n(\alpha_i,\Delta_i,k_i)$ transforms as a conformal correlator of operators with dimensions $\{\Delta_i\}$ on $S^d$. 


\subsection{Ambitwistor strings and the conformal basis}

Ambitwistor strings are worldsheet models that directly produce the CHY formulae \eqref{CHY0} as their worldsheet path-integral~\cite{Mason:2013sva,Adamo:2013tsa,Casali:2015vta}. Unlike ordinary string theories, they are chiral and constrained, containing only massless degrees of freedom in their spectrum. In general, the worldsheet action of an ambitwistor string theory takes the form (in conformal gauge)
\be\label{ats1}
S=\frac{1}{2\pi}\int_{\Sigma} P_{\mu}\,\dbar X^{\mu} -\frac{e}{2}\,P^{2}+ \cdots\,,
\ee
where $\Sigma$ is a closed Riemann surface, $X^\mu$ is a map from $\Sigma$ to the target space-time, and $P_{\mu}$ has conformal weight $(1,0)$ on $\Sigma$ (i.e., a section of the holomorphic canonical bundle $K_\Sigma$). The constraint $P^2=0$ is enforced through the Lagrange multiplier field $e$, which is a Beltrami differential on $\Sigma$ of conformal weight $(-1,1)$. The additional terms `$+\cdots$' are matter content, determining which particular space-time theory is described by the ambitwistor string.

Consider the matter content of: $d+2$ left-moving fermionic spinors and a left-moving worldsheet current algebra for the Lie algebra $\mathfrak{g}$. This model is known as the heterotic ambitwistor string, with worldsheet action:
\be\label{het1}
S=\frac{1}{2\pi}\int_{\Sigma} P_{\mu}\,\dbar X^{\mu} -\frac{e}{2}\,P^{2}+ \frac{1}{2} \Psi_{\mu}\,\dbar\Psi^{\mu}-\chi\,\Psi\cdot P + S_{\mathfrak{g}}\,,
\ee
where $S_{\mathfrak{g}}$ is the worldsheet action for the current algebra and $\chi$ is a fermionic Lagrange multiplier of conformal weight $(-\frac{1}{2},1)$ enforcing the constraint $\Psi\cdot P=0$. The action \eqref{het1} is invariant with respect to holomorphic conformal transformations on $\Sigma$, as well as a bosonic gauge freedom associated to the constraint $P^2=0$ and a fermionic gauge freedom associated to the constraint $\Psi\cdot P=0$. Fixing these redundancies leads to a gauge-fixed action
\be\label{het2}
S=\frac{1}{2\pi}\int_{\Sigma} P_{\mu}\,\dbar X^{\mu} + \frac{1}{2} \Psi_{\mu}\,\dbar\Psi^{\mu}+ b\,\dbar c+\tilde{b}\,\dbar\tilde{c}+\beta\,\dbar\gamma+S_{\mathfrak{g}}\,,
\ee
and associated BRST charge
\be\label{hetBRST}
Q=\oint cT+bc\partial c+\frac{\tilde{c}}{2}\,P^2+\gamma\,\Psi\cdot P+\frac{\tilde{b}}{2}\,\gamma^2\,,
\ee
where ghost systems $(b,c)$, $(\tilde{b},\tilde{c})$ and $(\beta,\gamma)$ are associated with holomorphic conformal transformations, the (bosonic) gauge freedom coming from $P^2=0$ and the (fermionic) gauge freedom coming from $\Psi\cdot P=0$, respectively. $T$ is the holomorphic stress tensor including all current algebra contributions. 

Vertex operators in the ambitwistor string CFT live in the BRST cohomology associated to $Q$. Consider a fixed vertex operator of the form:
\be\label{bavo0}
V=c\tilde{c}\,\delta(\gamma)\,j^{\sa}\,\Psi^{\mu}\,A^{\sa}_{\mu}(X)\,,
\ee
where $\sa$ is an adjoint index of $\mathfrak{g}$ and $j^{\sa}$ is the associated worldsheet current of conformal weight $(1,0)$. It is easy to see that the condition $QV=0$ imposes the Maxwell equation and Lorenz gauge condition on $A^{\sa}_{\mu}$. So \emph{any} basis of gluon wavefunctions can be used to construct $Q$-closed vertex operators, including the conformal basis.

In particular, choose $A^{\sa}_{\mu}(X)=\mathsf{T}^{\sa}\,A^{\Delta,\pm}_{\mu}(X;k)$ from \eqref{glCB2}, where $\mathsf{T}^{\sa}$ are generators of $\mathfrak{g}$. This gives the fixed vertex operator
\be\label{hetfvo}
\begin{split}
V^{-1} & = c\tilde{c}\,\delta(\gamma)\,j\cdot\mathsf{T}\,\Psi^{\mu}\int_0^\infty \frac{\d t}{t}\,t^{\Delta}\,\eps_\mu\, \e^{\pm\im tk\cdot X-\veps t} \\
  & = (\mp\im)^\Delta\,\Gamma(\Delta)\,c\tilde{c}\,\delta(\gamma)\,\frac{j\cdot\mathsf{T}\,\Psi\cdot\epsilon}{(-k\cdot X\mp\im\veps)^\Delta}\,,
\end{split}  
\ee
where the superscript on $V^{-1}$ denotes its picture number. The vertex operator in zero picture number is given by convolving the operator obtained from the descent procedure on momentum eigenstates with the conformal primary wavefunction:
\be\label{hetvo2}
V^{0} =c\tilde{c}\,j\cdot\mathsf{T}\,\int_0^\infty\frac{\d t}{t}\,t^{\Delta}\left(\eps\cdot P\pm \im t\eps\cdot\Psi\,k\cdot\Psi\right)\e^{\pm\im tk\cdot X-\veps t}\,,
\ee
and the integrated vertex operator is
\be\label{hetivo}
U=\int_{0}^{\infty}\frac{\d t}{t}\,t^{\Delta}\int_{\Sigma} \,j\cdot\mathsf{T}\,\left(\eps\cdot P\pm \im t\eps\cdot\Psi\,k\cdot\Psi\right)\e^{\pm\im tk\cdot X-\veps t}\,\bar\delta(tk\cdot P)\,,
\ee
 The holomorphic delta function $\bar\delta(z)=\bar\partial z^{-1}= \delta^2(z) \d\bar z$, so $\bar \delta(tk\cdot P)$ is understood to have support where $k\cdot P$  vanishes.

\smallskip

The conformal basis of graviton modes can be realized through vertex operators of the type II ambitwistor string, for which the matter content are two sets of left-moving fermionic spinors on the worldsheet. After fixing holomorphic worldsheet conformal transformations in addition to the various symmetries generated by constraints, one obtains the gauge-fixed action
\be\label{wsaII}
S=\frac{1}{2\pi}\int_{\Sigma} P_{\mu}\,\dbar X^{\mu} + \frac{1}{2} \Psi_{\mu}\,\dbar\Psi^{\mu}+\frac{1}{2}\tilde{\Psi}_{\mu}\,\dbar\tilde{\Psi}^{\mu}+ b\,\dbar c+\tilde{b}\,\dbar\tilde{c}+\beta\,\dbar\gamma+\tilde{\beta}\,\dbar\tilde{\gamma}\,,
\ee
and BRST charge
\be\label{IIBRST}
Q= \oint cT+bc\partial c+\frac{\tilde{c}}{2}\,P^2+\gamma\,\Psi\cdot P+\tilde{\gamma}\,\tilde{\Psi}\cdot P+\frac{\tilde{b}}{2}\left(\gamma^2+\tilde{\gamma}^2\right)\,.
\ee
Vertex operators in the BRST cohomology encode the linearised Einstein equations in de Donder gauge (together with a $B$-field and dilaton if desired), and once again we are free to construct explicit realizations using any basis of solutions, including the conformal basis.

In the conformal basis, the fixed graviton vertex operator is given by:
\be\label{IIfvo}
V^{-1,-1} = (\mp\im)^\Delta\,\Gamma(\Delta)\, c\tilde{c}\,\delta(\gamma)\,\delta(\tilde{\gamma})\,\frac{\eps\cdot\Psi\,\tilde\eps\cdot\tilde\Psi}{(-k\cdot X\mp\im\veps)^{\Delta}}\,,
\ee
using the representative \eqref{grCB2}. The picture number zero and integrated graviton vertex operators are
\be\label{IIvo2}
V^{0,0}=c\tilde{c}\,\int_0^\infty\frac{\d t}{t}t^{\Delta}\bigl(\eps\cdot P\pm\im t\eps\cdot\Psi\,k\cdot\Psi\bigr)\bigl(\tilde\eps\cdot P\pm\im t\tilde\eps\cdot\tilde\Psi\,k\cdot\tilde\Psi\bigr)\e^{\pm\im tk\cdot X-\veps t}\,,
\ee
and
\be\label{IIivo}
U=\int_0^\infty\frac{\d t}{t}\,t^{\Delta}\int_{\Sigma}\bigl(\eps\cdot P\pm \im t\eps\cdot\Psi\,k\cdot\Psi\bigr)\bigl(\tilde\eps\cdot P\pm\im t\tilde\eps\cdot\tilde\Psi\,k\cdot\tilde\Psi\bigr)\e^{\pm\im tk\cdot X-\veps t}\,\bar{\delta}(tk\cdot P)\,,
\ee
respectively.

\medskip

While we have only discussed gluon and graviton vertex operators here, it is easy to see that the conformal basis can be encoded in the vertex operators of any ambitwistor string (e.g., biadjoint cubic scalar theory, Einstein-Maxwell theory).


\subsection{CHY formulae}

The CHY formulae for tree-level scattering amplitudes in the plane wave basis are obtained from the genus zero (i.e., $\Sigma\cong\CP^1$) correlation functions of ambitwistor strings. With the vertex operators listed above, these correlation functions are easily computed in the conformal basis. First, consider the $n$-point worldsheet correlator of gluon vertex operators on $\Sigma\cong\CP^1$ in the heterotic ambitwistor string. Zero-mode saturation of the various worldsheet fields dictates that this correlation function is given by:
\be\label{gluon1}
A_{n}(\alpha_i,\Delta_i,k_i,\epsilon_i,\mathsf{T}^{\sa_i})=\left\la V_{1}^{-1}\,V_{2}^{-1}\,V_{3}^{0}\,\prod_{j=4}^{n}U_{j}\right\ra\,,
\ee
where we have explicitly listed the dependence of the amplitude on the various quantum numbers of the gluon conformal basis. 

Performing the $X$ path integral in \eqref{gluon1} fixes the worldsheet field $P_\mu$ to a classical value,
\be\label{Pform}
    P_\mu(\sigma) = \d \sigma\,\sum_{i=1}^n\frac{\alpha_i\,t_i\,k_{i\,\mu}}{\sigma-\sigma_i}\,,
\ee
where $\sigma$ is an affine coordinate on $\Sigma\cong\CP^1$ and $\{\sigma_i\}$ are the locations of the vertex operator insertions. The remaining portions of the path integral can also be performed explicitly, resulting in the celestial CHY formula:
\begin{multline}\label{gaugechy}
        A_n(\alpha_i,\Delta_i,k_i,\eps_i) = \prod_{i=1}^n\int_0^\infty\frac{\d t_i}{t_i}\,t_i^{\Delta_i}\,\e^{-\veps t_i}\,\delta^{d+2}\!\left(\sum_{j=1}^n \alpha_j\,t_j k_j\right) \\
        \times\int \frac{\sigma^{2}_{12}\sigma^{2}_{23}\sigma^{2}_{31}}{\d \sigma^{2}_1\,\d \sigma^2_2\,\d \sigma^2_3}\,\text{PT}_{n}\,\text{Pf}'M(\sigma, \alpha tk,\eps)\,\prod_{i=4}^{n}\bar\delta\!\left(\d \sigma_{i}\sum_{j\neq i}\frac{\alpha_i\alpha_jt_it_j\,k_i\cdot k_{j}}{\sigma_{ij}}\right)\,.
\end{multline}
Here $\sigma_{ij}:=\sigma_i-\sigma_j$; PT$_n$ is the worldsheet Parke-Taylor factor
\be\label{PT}
    \text{PT}_{n}:= \text{tr}(\mathsf{T}^{\sa_1}\mathsf{T}^{\sa_2}\cdots\mathsf{T}^{\sa_n})\frac{\d \sigma_1\,\d \sigma_2\,\cdots\,\d \sigma_n}{\sigma_{12}\,\sigma_{23}\cdots \sigma_{n1}} + \text{perms.}\,,
\ee
and Pf$'M$ is the reduced Pfaffian
\be\label{redpf}
\text{Pf}'M(\sigma, \alpha tk,\eps):=\frac{\sqrt{\d \sigma_1\,\d \sigma_2}}{\sigma_{12}}\,\text{Pf} M^{12}_{12}\,,
\ee
for $M^{12}_{12}$, the matrix $M$ below with rows and columns $1,2$ removed. This matrix $M$ is a slight modification of the usual CHY matrix,
\be\label{Matrix}
M(\sigma, \alpha tk, \eps):=\left( \begin{array}{cc}
                                A & -C^{\text{T}} \\
                                C & B
                               \end{array}\right)\,,
\ee
with entries
\be\label{Ments}
A_{ij}=\alpha_i\alpha_jt_i t_j\,k_i\cdot k_j\,\frac{\sqrt{\d \sigma_i\,\d \sigma_j}}{\sigma_{ij}}\,, \qquad B_{ij}=\eps_{i}\cdot\eps_j\,\frac{\sqrt{\d \sigma_i\,\d \sigma_j}}{\sigma_{ij}}\,,
\ee
\begin{equation*}
 C_{ij}=\alpha_jt_j\,\eps_i\cdot k_j\,\frac{\sqrt{\d \sigma_i\,\d \sigma_j}}{\sigma_{ij}}\,, \qquad C_{ii}=-\d \sigma_i\, \sum_{j\neq i}\alpha_jt_j\,\frac{\eps_i\cdot k_j}{\sigma_{ij}}\,.
\end{equation*}
It is the standard matrix appearing in the momentum space CHY formulae, but with the on-shell momenta replaced by $p_{i}\rightarrow \alpha_it_ik_i$.

It is straightforward to confirm that \eqref{gaugechy} is M\"obius invariant under transformations of the $\{\sigma_i\}$, and is furthermore permutation invariant. These properties follow from the $(d+2)$-dimensional delta function appearing under the Schwinger parameter integrals, which plays the role of overall momentum conservation in the plane wave basis. Indeed, \eqref{gaugechy} can be written more invariantly as:
\be\label{glchy}
\begin{split}
A_{n} = & \int \d\mu_{n}\,\text{PT}_{n}\,\prod_{i=1}^n\int_0^\infty\frac{\d t_i}{t_i}\,t_i^{\Delta_i}\,\e^{-\veps t_i} \\
        &\times \,\text{Pf}'M(\sigma, \alpha tk,\eps)\,\delta^{d+2}\!\left(\sum_{j=1}^n \alpha_j\,t_j k_j\right)\sideset{}{'}\prod_{j}\bar\delta\!\left(\sum_{l\neq j}\frac{\alpha_j\alpha_lt_jt_l\,k_j\cdot k_{l}}{\sigma_{jl}}\right)\,,
\end{split}
\ee
where the prime on the product in the second line indicates that only a M\"obius-invariant product of $n-3$ of these modified scattering equations are imposed. Note that \eqref{glchy} transforms as a conformal correlator on $S^d$ of spin-1 primaries with scaling dimensions $\{\Delta_i\}$; as expected, the difference between the states \eqref{glCB2} and the true conformal primary wavefunction \eqref{glCB1} drops out at the level of the gauge-invariant scattering amplitudes. 

\smallskip

For celestial graviton amplitudes, we perform the analogous worldsheet correlation function calculation in the type II ambitwistor string. At genus zero, the prescription for a $n$-point correlator is
\be\label{grav1}
A_{n}(\alpha_i,\Delta_i,k_i,\epsilon_i,\tilde{\epsilon}_i)=\left\la V_{1}^{-1,-1}\,V_{2}^{-1,-1}\,V_{3}^{0,0}\,\prod_{j=4}^{n}U_{j}\right\ra\,,
\ee
which leads to the CHY formula
\be\label{grchy}
\begin{split}
A_{n} = & \prod_{i=1}^n\int_0^\infty\frac{\d t_i}{t_i}\,t_i^{\Delta_i}\,\e^{-\veps t_i}\,\delta^{d+2}\!\left(\sum_{j=1}^n \alpha_j\,t_j k_j\right) \\
        &\times\int \d\mu_{n}\,\text{Pf}'M(\sigma,\alpha tk,\eps)\,\text{Pf}'M(\sigma,\alpha tk,\tilde{\eps})\,\sideset{}{'}\prod_{j}\bar\delta\!\left(\sum_{l\neq j}\frac{\alpha_j\alpha_lt_jt_l\,k_j\cdot k_{l}}{\sigma_{jl}}\right)\,.
\end{split}
\ee
Once again, the conformal basis analogues of overall momentum conservation and the scattering equations are universal, while the integrand is simply the familiar CHY integrand for gravitons with on-shell momenta replaced by $\alpha_it_ik_i$. It is easy to see that \eqref{grchy} transforms as a conformal correlator on $S^d$ of spin-2 primaries with scaling dimensions $\{\Delta_i\}$.

Similar reasoning for other ambitwistor string theories gives a general rule to obtain the CHY formula for celestial amplitudes:
\be\label{genchy}
\begin{split}
A_n(\alpha_i,\Delta_i,k_i,\ldots)= & \prod_{i=1}^n\int_0^\infty\frac{\d t_i}{t_i}\,t_i^{\Delta_i}\,\e^{-\veps t_i}\,\delta^{d+2}\!\left(\sum_{i=1}^n \alpha_i\,t_i k_i\right) \\
 & \times\int \d \mu_{n}\,\sideset{}{'}\prod_{j}\bar\delta\!\left(\sum_{l\neq j}\frac{\alpha_j\alpha_lt_jt_l\,k_j\cdot k_{l}}{\sigma_{jl}}\right)\, \mathcal{I}_{n}(\sigma, \alpha t k,\ldots)\,,
 \end{split}
\ee
where $\mathcal{I}_{n}(\sigma,\alpha t k,\ldots)$ is the CHY integrand relevant to the given QFT, with functional dependence on null momenta $p_i$ replaced by $\alpha_it_ik_i$.


\section{Conformal Soft Theorems}
\label{CSLs}

For certain values of the scaling dimension $\Delta$, conformal primary gluons or gravitons become pure gauge~\cite{Pasterski:2017kqt,Donnay:2018neh}. These special values are dimension independent: $\Delta=1$ for gluons and $\Delta=0,1$ for gravitons. For tree-level gluon scattering in $\R^{1,3}$, it was recently shown that in the limit where one of the gluons obeys $\Delta\rightarrow 1$, the celestial amplitude factorizes at leading order, with all dependence on this `conformally soft' gluon appearing in a conformal soft factor~\cite{Fan:2019emx,Pate:2019mfs,Nandan:2019jas}. This was confirmed by studying explicit examples of 4-dimensional gluon scattering in certain helicity configurations. Ambitwistor strings and the celestial CHY amplitudes allow us to probe conformal soft limits in arbitrary dimension and for \emph{all} tree-level gluon and graviton celestial amplitudes.


\subsection{Gauge theory: $\Delta\rightarrow 1$}

Consider the integrated gluon vertex operator given by \eqref{hetivo}. The Schwinger parameter integral can be performed explicitly to give an equivalent expression for this vertex operator:
\be\label{glivo}
U=(\mp\im)^{\Delta-1}\,\Gamma(\Delta-1)\int_{\Sigma}\frac{j\cdot\mathsf{T}\,\bar{\delta}(k\cdot P)}{(-k\cdot X\mp\im\veps)^{\Delta-1}}\left[\eps\cdot P+(\Delta-1)\frac{\eps\cdot\Psi\,k\cdot\Psi}{-k\cdot X\mp\im\veps}\right]\,.
\ee
This expression has a manifest pole at $\Delta=1$ due to the gamma function; a Laurent expansion of the vertex operator near $\Delta=1$ leads to
\be\label{gl11}
U=\frac{1}{\Delta-1}\int_{\Sigma}j\cdot\mathsf{T}\,\eps\cdot P\,\bar{\delta}(k\cdot P) + O((\Delta-1)^0):=\frac{U^{\mathrm{soft}}}{\Delta-1}+O((\Delta-1)^0)\,.
\ee
The leading conformally soft contribution to this vertex operator can be rewritten as
\be\label{gl1svo}
U^{\mathrm{soft}}=\frac{1}{2\pi\im}\oint \frac{j\cdot\mathsf{T}\,\eps\cdot P}{k\cdot P}\,,
\ee
where the holomorphic delta function has been expressed as a contour integral on the worldsheet containing the pole at $k\cdot P=0$.

When $U^{\mathrm{soft}}$ is inserted into a worldsheet correlation function it will develop poles coming from Wick contractions against the other vertex operators (which have generic, not conformally soft, values of $\Delta$), which are picked out by the contour integral. At the level of the worldsheet correlator:
\be\label{gl1sl}
\lim_{\Delta\rightarrow1}\left\la V_{1}^{-1}\,V_{2}^{-1}\,V_{3}^{0}\,\prod_{i=4}^{n}U_i\,U\right\ra = \frac{1}{\Delta-1}\left\la V_{1}^{-1}\,V_{2}^{-1}\,V_{3}^{0}\,\prod_{i=4}^{n}U_i\,U^{\mathrm{soft}}\right\ra\,,
\ee
where all sub-leading terms have been dropped on the right-hand side. Evaluating the contour integral of the conformally soft vertex operator by using \eqref{Pform} within this correlation function gives:
\begin{multline}\label{gl1sl2}
\frac{1}{\Delta-1}\sum_{\rho\in S_n/\Z_n}\sum_{j=1}^{n} \mathrm{tr}\left(\mathsf{T}^{\sa_{\rho(1)}}\cdots[\mathsf{T}^{\sa},\,\mathsf{T}^{\sa_{\rho(j)}}]\cdots \mathsf{T}^{\sa_{\rho(n)}}\right) \left\la \frac{\eps\cdot P(\sigma_{\rho(j)})}{k\cdot P(\sigma_{\rho(j)})}\,V_{1}^{-1}\,V_{2}^{-1}\,V_{3}^{0}\,\prod_{i=4}^{n}U_i\,\right\ra \\
= \frac{1}{\Delta-1}\,\sum_{\rho\in S_n/\Z_n}\sum_{j=1}^{n} \mathrm{tr}\left(\mathsf{T}^{\sa_{\rho(1)}}\cdots[\mathsf{T}^{\sa},\,\mathsf{T}^{\sa_{\rho(j)}}]\cdots \mathsf{T}^{\sa_{\rho(n)}}\right)\frac{\eps\cdot k_{\rho(j)}}{k\cdot k_{\rho(j)}}\left\la V_{1}^{-1}\,V_{2}^{-1}\,V_{3}^{0}\,\prod_{i=4}^{n}U_i\right\ra\,,
\end{multline}
where the colour structure of each state has been explicitly displayed and the sum is over distinct colour orderings. The formula follows from observing that the contraction of $P$ with  an exponential $e^{ik\cdot X}$  yields $k$.

The remainder of the correlator can be evaluated as before. Using the equivalence between the ambitwistor string correlators and the celestial amplitudes, \eqref{gl1sl} gives a relation between scattering amplitudes in the $\Delta\rightarrow 1$ limit. For the colour ordering with the conformally soft gluon inserted between $n$ and $1$, this gives the conformal soft gluon theorem:
\be\label{gl1sl3}
\lim_{\Delta\rightarrow 1}A_{n+1}(\Delta_1,\ldots,\Delta_n,\Delta)=\frac{1}{\Delta-1}\,\left(\frac{\eps\cdot k_1}{k\cdot k_1}-\frac{\eps\cdot k_n}{k\cdot k_n}\right)\,A_{n}(\Delta_1,\ldots,\Delta_n) +\cdots\,,
\ee
where $\eps_{\mu}$ is the polarization of the conformally soft gluon, $k_\mu$ is its insertion point on $S^d$ and $\{k_i\}$ are the insertion points on $S^d$ of the remaining gluons which have generic scaling dimensions. The `$\cdots$' indicate subleading terms of order $(\Delta-1)^0$.

\medskip

The formula \eqref{gl1sl3} is valid in arbitrary dimension, but it can be refined for four-dimensional scattering on $\R^{1,3}$ (i.e., $d=2$). In the momentum eigenstate basis, gluon polarizations can be traded for a helicity label and on-shell 4-momenta are specified by two-component Weyl spinors of opposite chirality $p_{\mu}\leftrightarrow (\lambda_{\alpha},\tilde{\lambda}_{\dot\alpha})$. In the conformal basis, this on-shell 4-momentum is traded for a point $(z,\bar{z})\in S^2$ and a scaling dimension $\Delta$. At the level of spinors,
\be\label{4dkin}
\lambda_{\alpha} = \sqrt{t}\begin{pmatrix}-z\\ 1\end{pmatrix}\equiv\sqrt{t}z_{\alpha},\qquad\tilde\lambda_{\dot\alpha} = \pm\sqrt{t}\begin{pmatrix}\bar{z}\\-1\end{pmatrix}\equiv\pm\sqrt{t}\tilde{z}_{\dot\alpha}\,,
\ee
with $z_\alpha$ homogeneous coordinates on the sphere $S^2\cong\CP^1$.

Details of the $d=2$ formalism are in the appendix~\ref{4D}, but for a positive helicity conformally soft gluon one obtains the soft theorem
\be\label{4dglst}
\begin{split}
\lim_{\Delta\rightarrow1}A_{n+1}(\Delta^{\pm}_1,\ldots,\Delta^{\pm}_{n},\Delta^{+}) & =\frac{-1}{\Delta-1}\,\frac{\la 1\,n\ra}{\la1\,s\ra\,\la s\,n\ra}\, A_{n}(\Delta^{\pm}_1,\ldots,\Delta^{\pm}_{n})+\cdots \\
& = \frac{-1}{\Delta-1}\,\frac{z_{1n}}{z_{1s}\,z_{sn}}\, A_{n}(\Delta^{\pm}_1,\ldots,\Delta^{\pm}_{n})+\cdots\,,
\end{split}
\ee
where $\Delta^{\pm}_{i}$ is short-hand for the scaling-dimension and helicity of each external gluon, $z_s\in S^2$ is the location of the $\Delta\rightarrow 1$ gluon, and we have expressed the right-hand side in both homogeneous and affine coordinates on $S^2$. In the former case, $\la i\,j\ra$ stands for the SL$(2,\C)$-invariant inner product $z_{i}^{\alpha}z_{j\,\alpha}=\epsilon^{\alpha\beta}z_{i\,\beta}z_{j\,\alpha}$. The formula \eqref{4dglst} agrees with the results of~\cite{Fan:2019emx,Pate:2019mfs,Nandan:2019jas}.


\subsection{Gravity: $\Delta\rightarrow 1$}

The integrated graviton vertex operator \eqref{IIivo} is equal to
\begin{multline}\label{grivo}
 U=(\mp\im)^{\Delta-1}\,\Gamma(\Delta-1)\int_{\Sigma}\frac{\bar{\delta}(k\cdot P)}{(-k\cdot X\mp\im\veps)^{\Delta-1}}\Bigg[\eps\cdot P\,\tilde\eps\cdot P  \\
 \left.+\frac{(\Delta-1)}{(-k\cdot X\mp\im\veps)}\left(\tilde\eps\cdot P\,\eps\cdot\Psi\,k\cdot\Psi+\eps\cdot P\,\tilde\eps\cdot\tilde\Psi\,k\cdot\tilde\Psi\right)+\Delta(\Delta-1)\,\frac{\eps\cdot\Psi\,k\cdot\Psi\,\tilde\eps\cdot\tilde\Psi\,k\cdot\tilde\Psi}{(-k\cdot X\mp\im\veps)^2}\right]\,,
\end{multline}
upon performing the Schwinger parameter integrals. As in the gluon case, the vertex operator has a simple pole as $\Delta\rightarrow 1$ due to the gamma function, and we can expand
\be\label{gr1sexp}
U=\frac{U^{\mathrm{soft}}}{\Delta-1}+O((\Delta-1)^0)\,,
\ee
with 
\be\label{gr1svo}
U^{\mathrm{soft}}=\frac{1}{2\pi\im}\oint \frac{\eps\cdot P\,\eps\cdot P}{k\cdot P}\,.
\ee
Here, we have explicitly symmetrized the graviton polarization by setting $\tilde\eps_\mu=\eps_\mu$, and the contour integral on the worldsheet $\Sigma$ surrounds $k\cdot P=0$. 

Now, inserting the leading conformally soft graviton operator into a worldsheet correlator of the type II ambitwistor string leads to:
\begin{multline}\label{gr1sl}
\frac{1}{\Delta-1}\left\la V_{1}^{-1,-1}\,V_{2}^{-1,-1}\,V_{3}^{0,0}\,\prod_{i=4}^{n}U_i\,U^{\mathrm{soft}}\right\ra \\ = \frac{1}{\Delta-1} \frac{1}{2\pi\im}\left\la\oint\,\frac{\eps\cdot P(\sigma)\,\eps\cdot P(\sigma)}{k\cdot P(\sigma)}\,V_{1}^{-1,-1}\,V_{2}^{-1,-1}\,V_{3}^{0,0}\,\prod_{i=4}^{n}U_i\right\ra\,,
\end{multline}
where the contour integral encircles poles in $\sigma\in\Sigma$, the insertion location of $U^{\mathrm{soft}}$. The remainder of the correlator can be evaluated, but there is a new subtlety which did not appear in the gluon calculation. Using \eqref{Pform}, it follows that
\be\label{gr1sl2}
\begin{split}
\frac{1}{2\pi\im}\oint\,\frac{\eps\cdot P(\sigma)\,\eps\cdot P(\sigma)}{k\cdot P(\sigma)} & =\frac{1}{2\pi\im}\oint \d \sigma \sum_{i=1}^{n}\alpha_i\,t_i\,\frac{\eps\cdot k_i\,\eps\cdot k_i}{k\cdot k_i\,(\sigma-\sigma_i)}+O((\sigma-\sigma_i)^0) \\
& = \sum_{i=1}^{n}\alpha_i\,t_i\,\frac{\eps\cdot k_i\,\eps\cdot k_i}{k\cdot k_i}\,.
\end{split}
\ee
The additional power of the Schwinger parameter $t_i$ means that the scaling dimensions of the remaining graviton states are shifted in the $\Delta\rightarrow 1$ limit.

Thus, we obtain the $\Delta\rightarrow 1$ conformal soft graviton theorem:
\begin{multline}\label{gr1sl3}
\lim_{\Delta\rightarrow 1}A_{n+1}(\Delta_1,\ldots,\Delta_n,\Delta) \\ =\frac{1}{\Delta-1}\,\sum_{i=1}^{n}\alpha_i\,\frac{\eps\cdot k_i\,\eps\cdot k_i}{k\cdot k_i}\,A_{n}(\Delta_1,\ldots,\Delta_i+1,\ldots,\Delta_n) +\cdots\,,
\end{multline}
where the scaling dimension of the $i^{\text{th}}$ external graviton is shifted to $\Delta_i\rightarrow\Delta_i+1$ in the $i^{\text{th}}$ term.  This shift ensures that the soft limit has the correct overall scaling dimension $\Delta_i$ in each $k_i$. This conformal soft limit can be refined in four space-time dimensions:
\begin{multline}\label{4dgrst}
\lim_{\Delta\rightarrow1}A_{n+1}(\Delta^{\pm}_1,\ldots,\Delta^{\pm}_{n},\Delta^{+}) \\
 =\frac{1}{\Delta-1}\,\sum_{i=1}^n\alpha_i\frac{[i\,s]\,\la\xi\,i\ra^2}{\la i\,s\ra\,\la\xi\,s\ra^2}\, A_{n}(\Delta^{\pm}_1,\ldots,(\Delta_i+1)^{\pm},\ldots,\Delta^{\pm}_{n})+\cdots \\
 = \frac{1}{\Delta-1}\,\sum_{i=1}^n\alpha_i\frac{\bar{z}_{is}\,(\xi-z_i)^2}{z_{is}\,(\xi-z_s)^2}\, A_{n}(\Delta^{\pm}_1,\ldots,(\Delta_i+1)^{\pm},\ldots,\Delta^{\pm}_{n})+\cdots\,,
\end{multline}
where $\xi\in S^2$ is an arbitrary reference point on the two-sphere\footnote{We thank Andrea Puhm~\cite{Puhm:2019zbl} for pointing out a typo in passing to the final line of this formula in an earlier version.}.


\subsection{Gravity: $\Delta\rightarrow 0$}

In~\cite{Pasterski:2017kqt}, it was shown that graviton conformal primaries with $\Delta=0$ are also pure gauge. Sure enough, the graviton vertex operator \eqref{grivo} also has a simple pole at $\Delta=0$ due to the gamma function. The vertex operator can be expanded around the $\Delta\rightarrow 0$ limit as
\be\label{gr0sl1}
U=\frac{U^{\mathrm{soft}}}{\Delta}+O(\Delta^0)\,,
\ee
where the new soft vertex operator is given by
\be\label{gr0svo}
\begin{split}
U^{\mathrm{soft}} & = \pm\frac{1}{2\pi}\oint\frac{\eps\cdot P}{k\cdot P}\left(\eps\cdot P\,k\cdot X+\eps\cdot\tilde\Psi\,k\cdot\tilde\Psi+\eps\cdot\Psi\,k\cdot\Psi\right) \\
&=\pm\frac{1}{2\pi}\oint \frac{\eps\cdot P}{k\cdot P}\,\eps^{\mu}\,k^{\nu}\left(P_{[\mu}\,X_{\nu]}+\Psi_{\mu}\,\Psi_{\nu}+\tilde\Psi_{\mu}\,\tilde\Psi_{\nu}\right)\,.
\end{split}
\ee
Once again, we have explicitly symmetrized the graviton polarization by setting $\tilde\eps_{\mu}=\eps_{\mu}$. In going from the first to second line of \eqref{gr0svo}, we added a term regular in $k\cdot P$ which does not contribute inside of correlation functions since there is no pole for the contour to wrap.

Inserting $U^{\text{soft}}$ in a correlation function of graviton vertex operators, one obtains a $\Delta\rightarrow 0$ conformal soft theorem for gravity by following steps virtually equivalent to the previous calculations:
\be\label{gr0sl}
\lim_{\Delta\rightarrow 0}A_{n+1}(\Delta_1,\ldots,\Delta_n,\Delta) =\frac{1}{\Delta}\,\sum_{i=1}^{n}\frac{\eps\cdot k_i}{k\cdot k_i}\,\eps_{\mu}\,k_{\nu}\,J_{i}^{\mu\nu}\,A_{n}(\Delta_1,\ldots,\Delta_n) +\cdots\,,
\ee
where the operator 
\be\label{Jop}
J_{i}^{\mu\nu}= 2\,\eps_i^{[\mu}\,\frac{\partial}{\partial\eps_{i\,\nu]}}+k_i^{[\mu}\,\frac{\partial}{\partial k_{i\,\nu]}}\,,
\ee
acts on the insertion points of the $i^{\mathrm{th}}$ graviton in $A_n$. Note the absence of any shifts in the scaling dimensions, as every $J_i^{\mu\nu}$ is homogenous in $k_i$. In four-dimensions, this conformal soft theorem takes the form\footnote{We thank Alfredo Guevara~\cite{Guevara:2019ypd} for pointing out an error in this formula in an earlier version.} 
\be\label{4dgrst2}
\begin{split}
\lim_{\Delta\rightarrow 0} A_{n+1}(\Delta^{\pm}_1,\ldots,\Delta^{\pm}_{n},\Delta^{+}) = \frac{1}{\Delta}\sum_{i=1}^{n}\frac{\bar{z}_{is}\,(\xi-z_i)}{z_{is}\,(\xi-z_s)}\left(\bar{z}_{si}\,\frac{\partial}{\partial\bar{z}_i}-2\bar h_i\right) A_{n}(\Delta^{\pm}_1,\ldots,\Delta^{\pm}_{n})\,,
\end{split}
\ee
where $\xi\in S^2$ is once again arbitrary, and $(h_i,\bar h_i) = \frac{1}{2}(\Delta_i+J_i,\Delta_i-J_i)$ represent the conformal weights of the gravitons. We find that the $\Delta\to 0$ conformal soft limit is just the Mellin transform of the subleading energetic soft limit of momentum space graviton amplitudes.


\section{Asymptotic Symmetries at $\scri$}
\label{asymptotics}

The relationship between energetic soft limits and asymptotic symmetries of gauge theory and gravity is well-established~(cf., \cite{He:2014laa,Lysov:2014csa,Strominger:2017zoo}). A similar relationship has been suggested between conformally soft wavefunctions and asymptotic symmetries~\cite{Donnay:2018neh,Fan:2019emx}, although the origins of this connection seem more obscure. By writing ambitwistor strings in variables explicitly adapted to the null conformal boundary $\scri$ of $\R^{1,d+1}$, we prove that the conformally soft vertex operators for gluons and gravitons are equivalent to charges which generate asymptotic symmetries of gauge theory and gravity.


\subsection{Ambitwistor strings at $\scri$}

The conformal boundary of $\R^{1,d+1}$ is composed of space-like, future and past time-like and future and past null infinities. The null boundaries $\scri^{\pm}$ are the endpoints of null trajectories followed by massless particles during a scattering process. Recall that $\scri^{\pm}\cong \R\times S^{d}$. 

For ambitwistor strings we work in complexified space-time $\C^{d+2}$ for which the associated $\scri_{\C}$ has the topology of a line bundle of Chern class one  $\cO(1)$  over a projective quadric,
the complexified $d$-sphere. Let $\zeta_{\mu}$ be homogeneous coordinates obeying $\zeta^2=0$; these give homogeneous coordinates on $S^d_{\C}$. Then $\scri_\C$ is described by coordinates $(u,\zeta_{\mu})$ subject to the homogeneity relation  
\begin{equation}\label{homogeneity}
    (u,\,\zeta_\mu)\sim(ru,\,r\zeta_\mu)\,,\qquad \forall r\in\C^*\,.
\end{equation}
If $(w,q^{\mu})$ are the conjugate (dual) coordinates to $(u, \zeta_{\mu})$, then $(u,\zeta_\mu,w,q^\mu)$ chart $T^*\scri_\C$, which is the space of all (complexified) null geodesics.

It is easy to see that this space is precisely the target space of ambitwistor string theory. Consider the generic model for such a worldsheet theory \eqref{ats1}, which is written in terms of canonical coordinates $(X^{\mu},P_{\mu})$ on $T^*\C^{d+2}$. The constraint $P^2=0$ reduces $T^*\C^{d+2}$ to the bundle of null directions, and its associated gauge freedom (which allows translations $X^{\mu}\rightarrow X^{\mu}+r\,P^{\mu}$) further quotients by motion along these null directions. The coordinates $(X^{\mu},P_{\mu})$ can be related explicitly to those on $T^*\scri_\C$ by
\begin{equation}\label{scritransform}
    X^\mu = w^{-1}\,q^\mu\,,\qquad P_\mu = w\,\zeta_\mu,\qquad w\,u = X\cdot P = q\cdot\zeta\,,
\end{equation}
with the variable $w$ acting like a frequency for the field $P_\mu$.

The generic worldsheet action \eqref{ats1} of ambitwistor string theory can now be rewritten with the worldsheet fields given by coordinates on $T^*\scri_\C$~\cite{Geyer:2014lca}. With respect to the symplectic structure
\begin{equation}
    \theta = w\,\d u - q^\mu\,\d\zeta_\mu\,,
\end{equation}
the homogeneity relation \eqref{homogeneity} has as its canonical generator the constraint,
\begin{equation}
    wu-q\cdot\zeta = 0\,.
\end{equation}
Imposing this constraint along with $\zeta^2=0$ at the level of a worldsheet action leads to the generic form of an ambitwistor string written in $\scri$-adapted variables~\cite{Geyer:2014lca}:
\be\label{atsscri}
    S = \frac{1}{2\pi}\int_\Sigma w\,\dbar u-q^\mu\,\dbar\zeta_\mu - \frac{e}{2}\,\zeta^2 - a\,(wu-q\cdot\zeta)+\cdots\,.
\ee
Here, $e$ and $a$ are Lagrange multipliers on the worldsheet which enforce the constraints, and `$+\cdots$' stands for whatever matter content is chosen for the particular realization of the ambitwistor string.


\subsection{Conformally soft vertex operators as asymptotic symmetry generators}

Conformal primary vertex operators for gluons and gravitons can be written down, and the CHY formulae \eqref{gaugechy}, \eqref{grchy} derived using the $\scri$-adapted ambitwistor strings. However, the real power of this perspective is in allowing a geometric interpretation for the action of worldsheet operators at $\scri$. Consider the vertex operator for a $\Delta=1$ conformally soft gluon \eqref{gl1svo}; written in the $\scri$ variables it takes the form:
\be\label{scrigl1}
U^{\Delta=1}_{\text{gluon}}=\frac{1}{2\pi\im}\oint j\cdot\mathsf{T}\,\frac{\eps\cdot\zeta}{k\cdot\zeta}\,.
\ee
This has the structure of a canonical charge generating some canonical transformations\footnote{In worldsheet theories, a canonical transformation corresponding to a Hamiltonian function $h(X,P)$ is generated via a charge $Q_h$ acting on functions $f(X,P)$ as
\begin{equation*}
    [Q_h,f(X,P)(z)] = \oint\frac{\d z'}{2\pi\im}\mathcal{R}[h(X,P)(z')\,f(X,P)(z)]\,,
\end{equation*}
with the contour of integration encircling poles at $z'=z$ and $\mathcal{R}$ denoting radial ordering in the 2d CFT.} in the heterotic ambitwistor string. Acting at $\scri$, this charge is the generator of a Kac-Moody algebra (through the worldsheet current $j^{\sa}$) corresponding to a gauge transformation on the celestial sphere $S^d$ with gauge parameter $\eps\cdot\zeta/k\cdot\zeta$~\cite{Geyer:2014lca}. Such large gauge transformations on $S^d$ are well-known asymptotic symmetries of Yang-Mills theory~\cite{Strominger:2013lka}.

A similar story holds for conformally soft graviton vertex operators. Translating into $\scri$-adapted variables, the $\Delta=1$ and $\Delta=0$ conformally soft graviton vertex operators are:
\be\label{scrigr1}
U^{\Delta=1}_\text{grav} = \frac{1}{2\pi\im}\oint w\,\frac{(\eps\cdot \zeta)^2}{k\cdot\zeta}\,,
\ee
and
\be\label{scrigr0}
U^{\Delta=0}_\text{grav}=\pm\frac{1}{2\pi}\oint\frac{\eps\cdot\zeta}{k\cdot\zeta}\,\eps^\mu\, k^\nu\,\left(\zeta_{[\mu}\,q_{\nu]} + \Psi_\mu\,\Psi_\nu + \tilde\Psi_\mu\,\tilde\Psi_\nu\right)\,.
\ee
As $w$ is the canonical conjugate of $u$, the coordinate on the generators of $\scri$, $U^{\Delta=1}_\text{grav}$ acts to shift $u$ by $\frac{(\eps\cdot \zeta)^2}{k\cdot\zeta}$. This is precisely the action of a supertranslation within the BMS group~\cite{Bondi:1962px,Sachs:1962wk}, though when $d>2$ these supertranslations are not naturally asymptotic symmetries (cf., \cite{Hollands:2003ie,Hollands:2016oma}).

In \eqref{scrigr0}, the terms in parentheses act on $\scri$ by the operator
\be\label{scriangmom}
J_{\mu\nu} = \zeta_{[\mu}\,q_{\nu]} + \Psi_{\mu}\,\Psi_{\nu} + \tilde\Psi_{\mu}\,\tilde\Psi_{\nu}\,,
\ee
which is a sum of orbital and intrinsic angular momentum operators~\cite{Geyer:2014lca}. Therefore, $U^{\Delta=0}_\text{grav}$ acts as a superrotation~\cite{Barnich:2009se,Barnich:2010eb} by $\frac{\eps\cdot\zeta}{k\cdot\zeta}\,\eps^\mu k^\nu$ on $\scri$.


\subsection{The $\Delta\rightarrow 0$ limit for subleading soft gluons}

The gluon vertex operator \eqref{glivo} also has a simple pole for scaling dimension $\Delta=0$; however, the gluon conformal primary wavefunction is \emph{not} pure gauge when $\Delta=0$. It therefore seems that the $\Delta\rightarrow 0$ limit for gluon conformal primary wavefunctions should have no special meaning. However, isolating the leading part of \eqref{glivo} gives
\be\label{scrigl0}
U^{\Delta=0}_{\text{gluon}}=\pm\frac{1}{2\pi}\oint \frac{j\cdot \mathsf{T}}{w\,k\cdot\zeta}\left(\eps\cdot\zeta\, k\cdot q + \eps\cdot\Psi\, k\cdot\Psi\right)\,.
\ee
This acts on $\scri$ through a mixture of gauge transformation and rotation, which is precisely the Yang-Mills analogue of a superrotation~\cite{Lysov:2014csa,Himwich:2019dug}. This is perhaps puzzling, since it appears to link an asymptotic symmetry of gauge theory with a non-soft limit of the scaling dimension.

Furthermore,  upon inserting \eqref{scrigl0} into a worldsheet correlation function, a universal decoupling behaviour akin to a conformal soft theorem emerges: 
\begin{multline}\label{gl0sl}
\lim_{\Delta\rightarrow 0}A_{n+1}(\Delta_1,\ldots,\Delta_n,\Delta) =\frac{\eps_{\mu}\,k_{\nu}}{\Delta}\,\left(\frac{J_{1}^{\mu\nu}}{k\cdot k_1}\,A_{n}(\Delta_1-1,\Delta_2,\ldots,\Delta_n)\right. \\
\left.-\frac{J_{n}^{\mu\nu}}{k\cdot k_{n}}\,A_{n}(\Delta_1,\ldots,\Delta_{n-1},\Delta_{n}-1)\right)+\cdots\,,
\end{multline}
where we have displayed the result for the partial amplitude with the $\Delta\rightarrow 0$ gluon inserted between $1$ and $n$ in the colour-ordering. Here $J_i^{\mu\nu}$ is angular momentum operator 
\be\label{Jop2}
J_{i}^{\mu\nu}=\eps_i^{[\mu}\,\frac{\partial}{\partial\eps_{i\,\nu]}}+k_i^{[\mu}\,\frac{\partial}{\partial k_{i\,\nu]}}\,,
\ee

\noindent{}for the $i^{\mathrm{th}}$ gluon. Note that in this limit, the reduced amplitudes appear with scaling dimensions shifted \emph{down}: $\Delta_{1,n}\rightarrow \Delta_{1,n}-1$. 

The structural form of \eqref{gl0sl} is clearly inherited from the sub-leading energetic soft gluon theorem. This follows directly from \eqref{glCB1} or \eqref{glCB2} by evaluating the conformal primary wavefunction at $\Delta=0$. This leads to a vertex operator at $\Delta=0$ that corresponds precisely to the subleading part of the standard momentum vertex operator. It would be very interesting to understand this from some underlying (perhaps holographic) framework.

More generally, it is clear that the conformal primary vertex operators for both gauge theory and gravity have simple poles for all negative integer values of $\Delta$. While these do not correspond to pure gauge configurations (like the $\Delta=0$ gluon example discussed here), one still expects the S-matrix to factorize in a way that is structurally related to the energetic soft expansion. For instance, the $\Delta\rightarrow-1$ limit for gravitons has been shown to be related to the sub-sub-leading soft graviton theorem using recursive techniques~\cite{Guevara:2019ypd}.

\acknowledgments

We thank Alfredo Guevara and Andrea Puhm for pointing out typos and errors~\cite{Puhm:2019zbl,Guevara:2019ypd} in an earlier version of this paper. TA is supported by an Imperial College Junior Research Fellowship. AS is supported by a Mathematical Institute Studentship, Oxford.


\appendix
\section{Results in Four-dimensions}
\label{4D}

The case of celestial scattering in $\R^{1,3}$ (i.e., $d=2$) is special for several reasons: scattering is refined by helicity instead of polarizations, the celestial sphere $S^2$ carries the full Virasoro algebra, and the conformally soft limit $\Delta\rightarrow1$ occurs within the conformal basis relevant for this dimension. In addition, the shadow transform of the graviton conformal primary with scaling dimension $2-\Delta$ becomes conformally soft as $\Delta\rightarrow 2$~\cite{Pasterski:2017kqt}. In this appendix, we give a brief derivation of conformal soft theorems with $d=2$, using four-dimensional ambitwistor strings~\cite{Geyer:2014fka}.


\subsection{Ambitwistor strings in 4-dimensions}

The space of (complexified) null geodesics in $\R^{1,3}$ is given by a projective quadric in $\CP^3\times\CP^3$. If each factor is endowed with homogeneous coordinates $Z^{A}=(\mu^{\dot\alpha},\lambda_{\alpha})$, $W_{A}=(\tilde{\lambda}_{\dot\alpha},\tilde{\mu}^\alpha)$, then this quadric is $Z\cdot W=[\mu\,\tilde{\lambda}]+\la\tilde{\mu}\,\lambda\ra=0$. A worldsheet theory adapted to this target space is given in conformal gauge by 
\be\label{4dats}
    S = \int_\Sigma Z\cdot\dbar W - W\cdot\dbar Z  + a\,Z\cdot W+\cdots\,,
\ee
where the worldsheet fields $(Z^A,W_A)$ are bosonic spinors on $\Sigma$ of conformal dimension $(\frac{1}{2},0)$ and $a$ is a Lagrange multiplier of conformal weight $(0,1)$ enforcing the constraint $Z\cdot W=0$. The `$+\cdots$' stand for theory-dependent matter content. 

In a plane wave basis, the data for an on-shell gluon or graviton is given by a helicity sign and an on-shell 4-momentum, represented by spinors $\{\lambda_{\alpha},\tilde{\lambda}_{\dot\alpha}\}$. To transform to the conformal basis, this data must be transformed to $\{\Delta,z,\bar{z}\}$, where $\Delta$ is the scaling dimension and $(z,\bar{z})\in S^2$. This is accomplished with the conventions
\begin{equation}\label{spinorconvention}
    \lambda_{\alpha} = \sqrt{t}\begin{pmatrix}-z\\ 1\end{pmatrix}\equiv\sqrt{t}z_{\alpha},\qquad\tilde\lambda_{\dot\alpha} = \pm\sqrt{t}\begin{pmatrix}\bar{z}\\-1\end{pmatrix}\equiv\pm\sqrt{t}\tilde{z}_{\dot\alpha}\,,
\end{equation}
where the sign in the second relation denotes whether the momentum is outgoing/incoming, and $z_{\alpha}$ define homogeneous coordinates on $S^2\cong\CP^1$. This enables vertex operators in models \eqref{4dats} to be obtained by a simple Mellin transform of those for plane wave momentum eigenstates.

For Yang-Mills theory, the matter content needed in \eqref{4dats} is simply a left-moving worldsheet current algebra for the gauge group~\cite{Geyer:2014fka}. Integrated vertex operators for positive and negative helicity gluons in the conformal basis are then given by:
\begin{equation}\label{ymvertex}
    \begin{split}
        &U = \int_\Sigma\int_0^\infty\frac{\d t}{t}\,t^{\Delta}\int_{\C^*}\frac{\d s}{s}\,\bar\delta^{2}\!\left(\sqrt{t}z-s\lambda(\sigma)\right)\,\e^{\pm\im \sqrt{t}s[\mu(\sigma)\tilde{z}]-\veps t}\,j\cdot\mathsf{T}\,,\\
        &\tilde U = \int_\Sigma\int_0^\infty\frac{\d t}{t}\,t^{\Delta}\int_{\C^*}\frac{\d s}{s}\,\bar\delta^{2}\!\left(\pm\sqrt{t}\tilde{z}-s\tilde\lambda(\sigma)\right)\,\e^{\im \sqrt{t}s\langle\tilde\mu(\sigma)z\rangle-\veps t}\,j\cdot\mathsf{T}\,,
    \end{split}
\end{equation}
respectively. These vertex operators can be further simplified by scaling $s\rightarrow\sqrt{t}s$ in order to perform the $t$ integral, making a pole at $\Delta=1$ manifest. The bulk-to-boundary propagator forms of these vertex operators are also produced naturally from the $t$ integral, illustrating the correspondence between \eqref{ymvertex} and \eqref{hetivo}. 

For gravity, the worldsheet action \eqref{4dats} is supplemented with a complex fermion system $(\rho^{A},\tilde{\rho}_{A})$ of conformal weight $(\frac{1}{2},0)$ as well as several new fermionic and bosonic constraints (see \cite{Geyer:2014fka}). Positive and negative helicity graviton vertex operators in the conformal basis are given by:
\begin{equation}\label{gravityvertex}
    \begin{split}
         U = \int_\Sigma(1+\rho\cdot\partial_Z\,\tilde\rho\cdot\partial_W)\int_0^\infty\frac{\d t}{t}\,t^{\Delta}\int_{\C^*} & \frac{\d s}{s^2}\,\bar\delta^{2}\!\left(\sqrt{t}z-s\lambda(\sigma)\right)  \\
         & \times\,[\tilde\lambda(\sigma)\tilde{z}]\,\e^{\pm\im\sqrt{t} s[\mu(\sigma)\tilde{z}]-\veps t}\,,\\
         \tilde U = \int_\Sigma(1+\rho\cdot\partial_Z\,\tilde\rho\cdot\partial_W)\int_0^\infty\frac{\d t}{t}\,t^{\Delta}\int_{\C^*} & \frac{\d s}{s^2}\bar\delta^{2}\!\left(\pm\sqrt{t}\tilde{z}-s\tilde\lambda(\sigma)\right) \\
         & \times\,\langle\lambda(\sigma)z\rangle\,\e^{\im \sqrt{t}s\langle\tilde\mu(\sigma)z\rangle-\veps t}\,.
    \end{split}
\end{equation}
Once again, these are obtained by a Mellin transform of the momentum eigenstate graviton vertex operators in 4d ambitwistor strings.


\subsection{Conformal soft theorems in 4d}

The conformal soft theorems are derived in the same fashion as in general dimension: by expanding the vertex operators around the soft scaling dimensions and then inserting the leading soft vertex operator into a worldsheet correlation function. First, consider the $\Delta\rightarrow1$ conformal soft limit for a gluon, which we take to be positive helicity without loss of generality. Both the $t$ and $s$ integrals in \eqref{ymvertex} can be performed to yield 
\be\label{4dglop}
    U = (-\im)^{\Delta-1}\,\Gamma(\Delta-1)\int_\Sigma j\cdot\mathsf{T}\,\left(\frac{\langle\xi\lambda(\sigma)\rangle}{\langle\xi z\rangle}\right)^{\Delta}\,\frac{\bar\delta(\langle z\lambda(\sigma)\rangle)}{(-[\mu(\sigma)\tilde{z}]-\im\veps)^{\Delta-1}}\,,
\end{equation}
where $\xi\in S^2$ is an arbitrary reference point. Clearly, this has a pole at $\Delta=1$; the coefficient of $(\Delta-1)^{-1}$ in a Laurent expansion of \eqref{4dglop} is the operator
\be\label{4gl1}
 U^{\Delta=1}_{\text{soft}} = \frac{1}{2\pi\im}\oint j\cdot\mathsf{T}\frac{\langle\xi\lambda(\sigma)\rangle}{\langle\xi z\rangle\,\la z\lambda(\sigma)\ra}\,,
\end{equation}
where the contour encircles the pole at $\la z\lambda\ra=0$. As shown in~\cite{Geyer:2014lca,Adamo:2015fwa}, this is a generator of asymptotic gauge transformations in $\R^{1,3}$. Inserting \eqref{4gl1} in a worldsheet correlation function (between $1$ and $n$ in the colour ordering) produces the positive helicity conformal soft gluon theorem:
\be\label{4gl1lim}
\begin{split}
\lim_{\Delta\rightarrow1}A_{n+1}(\Delta^{\pm}_1,\ldots,\Delta^{\pm}_{n},\Delta^{+}) & =\frac{-1}{\Delta-1}\,\frac{\la1\,n\ra}{\la1\,s\ra\,\la s\,n\ra}\, A_{n}(\Delta^{\pm}_1,\ldots,\Delta^{\pm}_{n})+\cdots \\
& = \frac{-1}{\Delta-1}\,\frac{z_{1n}}{z_{1s}\,z_{sn}}\, A_{n}(\Delta^{\pm}_1,\ldots,\Delta^{\pm}_{n})+\cdots\,,
\end{split}
\ee
where the $\Delta\rightarrow 1$ gluon is inserted at $(z_s,\bar{z}_s)\in S^2$ and superscripts on the scaling dimensions indicate the helicity of the external gluon. The result is displayed in both homogeneous and affine coordinates on the celestial $S^2$.

\medskip

Likewise, the positive helicity graviton vertex operator of \eqref{gravityvertex} can be written as:
\be\label{4dgrop}
 U = (-\im)^{\Delta-1}\,\Gamma(\Delta-1)\int_\Sigma(1+\rho\cdot\partial_Z\,\tilde\rho\cdot\partial_W)\,\left(\frac{\langle\xi\lambda(\sigma)\rangle}{\langle\xi z\rangle}\right)^{\Delta+1}\,\frac{\bar\delta(\langle z\lambda(\sigma)\rangle)\,[\tilde\lambda(\sigma)\tilde{z}]}{(-[\mu(\sigma)\tilde{z}]-\im\veps)^{\Delta-1}}\,.
\ee
This has poles at both $\Delta=1$ and $\Delta=0$. In the first case the coefficient of $(\Delta-1)^{-1}$ is given by
\be\label{4gr1}
U^{\Delta=1}_{\text{soft}}=\frac{1}{2\pi\im}\oint (1+\rho\cdot\partial_Z\,\tilde\rho\cdot\partial_W)\,\left(\frac{\langle\xi\lambda(\sigma)\rangle}{\langle\xi z\rangle}\right)^2\,\frac{[\tilde\lambda(\sigma)\tilde{z}]}{\langle z\lambda(\sigma)\rangle}\,,
\ee
with the contour taken around the pole at $\langle z\lambda\rangle=0$. This vertex operator can be shown to generate supertranslations at $\scri$~\cite{Geyer:2014lca}, and inserting it into a worldsheet correlation function leads to the positive helicity conformal soft graviton theorem:
\begin{multline}\label{4gr1lim}
\lim_{\Delta\rightarrow1}A_{n+1}(\Delta^{\pm}_1,\ldots,\Delta^{\pm}_{n},\Delta^{+}) \\
 =\frac{1}{\Delta-1}\,\sum_{i=1}^n\alpha_i\frac{[i\,s]\,\la\xi\,i\ra^2}{\la i\,s\ra\,\la\xi\,s\ra^2}\, A_{n}(\Delta^{\pm}_1,\ldots,(\Delta_i+1)^{\pm},\ldots,\Delta^{\pm}_{n})+\cdots \\
 = \frac{1}{\Delta-1}\,\sum_{i=1}^n\alpha_i\frac{\bar{z}_{is}\,(\xi-z_i)^2}{z_{is}\,(\xi-z_s)^2}\, A_{n}(\Delta^{\pm}_1,\ldots,(\Delta_i+1)^{\pm},\ldots,\Delta^{\pm}_{n})+\cdots\,,
\end{multline}
with $\xi\in S^2$ arbitrary.

Expanding the graviton vertex operator \eqref{4dgrop} near $\Delta\rightarrow 0$, the leading term is 
\be\label{4gr0}
U^{\Delta=0}_{\mathrm{soft}}=\frac{1}{2\pi}\oint (1+\rho\cdot\partial_Z\,\tilde\rho\cdot\partial_W)\,\frac{\la\xi\lambda(\sigma)\ra\,[\tilde{\lambda}(\sigma)\tilde{z}]\,[\mu(\sigma)\tilde{z}]}{\la\xi z\ra\,\la z\lambda(\sigma)\ra}\,,
\ee
where the contour once again encircles the $\langle z\lambda\rangle=0$ pole. This operator acts as a charge generating superrotations at $\scri$~\cite{Geyer:2014lca}, and it produces the positive helicity conformal soft graviton theorem:
\be\label{4gr0lim}
\begin{split}
\lim_{\Delta\rightarrow 0} A_{n+1}(\Delta^{\pm}_1,\ldots,\Delta^{\pm}_{n},\Delta^{+}) = \frac{1}{\Delta}\sum_{i=1}^{n}\frac{\bar{z}_{is}\,(\xi-z_i)}{z_{is}\,(\xi-z_s)}\left(\bar{z}_{si}\,\frac{\partial}{\partial\bar{z}_i}-2\bar h_i\right) A_{n}(\Delta^{\pm}_1,\ldots,\Delta^{\pm}_{n})\,,
\end{split}
\ee
with $\bar h_i$ the anti-holomorphic conformal weight of the gravitons. This $\Delta\rightarrow 0$ limit is also of interest thanks to its interaction with the shadow transform when $d=2$.

In general dimension, the conformal primary wavefunction $h_{\mu\nu}^{\Delta,\pm}$ is related to its shadow transform by 
\be\label{shadow}
\widetilde{h_{\mu\nu}^{\Delta,\pm}}(X;k)=(-X^2)^{\frac{d}{2}-\Delta}\,h_{\mu\nu}^{d-\Delta,\pm}(X;k)\,,
\ee
where tilde denotes the shadow transform (cf., \cite{SimmonsDuffin:2012uy}). When $d=2$ and $\Delta=2$, $\widetilde{h_{\mu\nu}^{2,\pm}}$ is pure gauge, being closely related to the $\Delta\rightarrow 0$ limit. This $\Delta= 2$ pure diffeomorphism state can be used to construct a stress tensor for a boundary CFT on $S^2$ whose insertion into a correlator leads to the sub-leading energetic soft graviton theorem~\cite{Cheung:2016iub,Kapec:2016jld}. In the context of celestial scattering, \eqref{4gr0lim} indicates that $\Delta\rightarrow2$ will be linked with a conformal soft theorem, inheriting the functional form of this sub-leading soft graviton theorem.

\bibliography{CBA}
\bibliographystyle{JHEP}

\end{document}